# Structure and Short-time Dynamics in Suspensions of Charged Silica Spheres in the entire Fluid Regime

J. Gapinski and A. Patkowski

*Faculty of Physics, A. Mickiewicz University, 61-614 Poznan, Poland*

and

A. J. Banchio

*CONICET and FaMAF, Universidad Nacional de Córdoba, Ciudad Universitaria, X5000HUA Córdoba, Argentina*

and

J. Buitenhuis, P. Holmqvist, M. P. Lettinga, G. Meier and G. Nägele

*Institut für Festkörperforschung, Forschungszentrum Jülich, D-52425 Jülich, Germany*

**Abstract**

We present an experimental study of short-time diffusion properties in fluid-like suspensions of monodisperse charge-stabilized silica spheres suspended in DMF. The static structure factor $S(q)$, the short-time diffusion function, $D(q)$, and the hydrodynamic function, $H(q)$, in these systems have been probed by combining X-ray photon correlation spectroscopy experiments with static small-angle X-ray scattering. Our experiments cover the full liquid-state part of the phase diagram, including deionized systems right at the liquid-solid phase boundary. We show that the dynamic data can be consistently described by the renormalized density fluctuation expansion theory of Beenakker and Mazur over a wide range of concentrations and ionic strengths. In accord with this theory and Stokesian dynamics computer simulations, the measured short-time properties cross over *monotonically*, with increasing salt content, from the bounding values of salt-free suspensions to those of neutral hard spheres. Moreover, we discuss an upper bound for the hydrodynamic function peak height of fluid systems based on the Hansen-Verlet freezing criterion.

Corresponding author:
*Jacek Gapinski, electronic mail: gapinski@amu.edu.pl*

# I. INTRODUCTION

Dispersions of charge-stabilized colloidal particles immersed in a low-molecular-weight solvent are ubiquitous in environmental and food industry, and in life sciences. There are many examples of such dispersions including viruses or proteins in water, paints, waste water, and model systems of strongly charged spherical latex or silica spheres. The exploration of dynamic properties in these colloidal model systems is interesting in its own right. In addition, it can provide insights into the transport properties of more complex colloidal or macromolecular particles relevant to industry and biology. The dynamics of model systems of charged spheres is the subject of ongoing experimental and theoretical research (see Refs[1,2,3,4]). The theoretical understanding of the dynamics in colloidal model systems is still incomplete, since complicated many-body interactions have to be considered.

At long distances, charged colloidal spheres interact electrostatically by an exponentially screened Coulomb repulsion. This reflects the increase in free energy originating from the overlap of electric double layers around the colloids formed by surface-released counterions and residual salt ions. Colloidal particles influence each other also through solvent-mediated hydrodynamic interactions (HI) which, in unconfined suspensions of mobile particles, are long-ranged with fluid velocity perturbations decaying asymptotically like $1/r$ with the interparticle distance $r$. Therefore, HI are in general influential on the suspension dynamics both in equilibrium and non-equilibrium situations.

In dynamic scattering experiments, one distinguishes between the colloidal short-time regime, where $\tau_B \ll t \ll \tau_I$, and the long-time regime where $t \gg \tau_I$. Here, $\tau_B$ is the momentum relaxation time characterizing the fast relaxation of the colloid particle momentum into the Maxwellian equilibrium. This characteristic time is of the same order of magnitude as the viscous relaxation time, $\tau_\eta$, that sets the time scale for which solvent inertia matters. For the correlation times $t \gg \tau_B \sim \tau_\eta$ commonly probed in scattering experiments on colloids, both the particles and the solvent move quasi-inertia free, which allows for a pure configuration-space description of their dynamics based on the many-particle Smoluchowski equation in combination with the creeping flow equation of stationary low-Reynolds number solvent flow[3, 4].

The interaction time, $\tau_I$, can be estimated by $a^2/D_0$, where $D_0 = k_{BT}/(6\pi\eta_0 a)$ is the translational diffusion coefficient at infinite dilution, $a$ is the hydrodynamic sphere radius, and $\eta_0$ is the solvent shear viscosity. It gives a rough estimate of the delay time for which (non-hydrodynamic) direct particle forces become important. For long times $t \gg \tau_I$, the motion of particles is diffusive, due to the cumulative effect of the fast random bombardment by the solvent molecules, and due to the interactions with other surrounding particles originating from significant changes in their configuration. Within the colloidal short-time regime, the particle configuration has changed so little that the slowing influence of the electro-steric interactions is not yet operative. However, the short-time dynamics is directly influenced by the solvent-mediated HI which, as a salient remnant of the solvent degrees of freedom, act quasi-instantaneously for times $t \gg \tau_\eta$, and indirectly by the direct forces through their effect on the equilibrium microstructure. Long-time diffusion quantities are in general smaller than the corresponding short-time quantities, since the dynamic caging of particles by neighboring ones, operative at longer times ($t \geq \tau_I$), has an overall slowing influence on the overdamped dynamics. Short-time properties, on the other hand, are not subject to the dynamic caging effect since the configuration of neighboring particles hardly changes for $t \ll \tau_I$. Therefore, short-time properties are more simply expressed in terms of equilibrium averages. As far as the peak height in $H(q)$ is concerned, our experimental results bridge the gap between hard-sphere-like systems and fully deionized systems of strongly charged colloidal particles. The majority of earlier work was concerned with systems with peak heights in $H(q)$ not far from the corresponding equal-concentration hard-sphere values[5, 6].

In the present work, we describe a detailed experimental study of static and short-time diffusion properties in suspensions of very monodisperse, charged silica spheres suspended in DMF. We use here DMF rather than water as the suspending solvent so that the problem of $CO_2$ adsorption is avoided, and the most interesting regime of very low salt concentrations can be explored. This has allowed us to study systematically the low-salt regime of strongly correlated, but still fluid-like ordered particles close to the freezing transition into a crystal phase. In some other work, strikingly low values of $H(q)$ have been reported to occur in this low-salt regime[7, 8, 9, 10], an observation which was interpreted initially in terms of hydrodynamic screening[11]. Using the synchrotron radiation facility at the ESRF in Grenoble, we have performed X-ray photon correlation spectroscopy experiments (XPCS) to explore very systematically the concentration and ionic-strength dependence of the static structure factor, $S(q)$, and of short-time diffusion properties including the wavenumber-dependent function, $D(q)$, and the related hydrodynamic function $H(q)$. Our study encompasses the liquid phase regime, and explores in particular the fluid regime close to the liquid-solid phase boundary.

We show that all our experimental short-time diffusion data, obtained for three different series of samples, can be consistently described, and explained, by means of the renormalized density fluctuation expansion method of Beenakker and Mazur[12, 13, 14], commonly referred to as the δγ-scheme. This method includes many-body HI in an approximate way through the consideration of so-called ring diagrams. It is the most comprehensive analytical scheme available so far to calculate short-time dynamic properties. Originally, it had been applied to hard spheres only, but in later work it's zeroth-order version was used also to make predictions for charge-stabilized systems[3, 7, 11, 15, 16, 17, 18, 19]. This method requires the static structure factor as the basic input, which we calculate using the Rogers-Young (RY) and rescaled mean-spherical approximation (RMSA) integral equation schemes[20, 21]. The colloid pair interactions are described in these calculations on the basis of the one-component macroion fluid (OMF) model of dressed spherical macroions that interact by an effective screened Coulomb potential of Derjaguin-Landau-Verwey-Overbeek (DLVO) type. In recent accelerated Stokesian Dynamics (ASD) computer simulation studies based on the OMF model, it was shown that the δγ-scheme allows for the prediction of short-time diffusion properties on a semi-quantitative level of accuracy[22, 23, 24, 25, 26].

In this paper, we have achieved a thorough experimental test of the δγ-scheme by measuring simultaneously the short-time diffusion function and the static structure for each individual sample. For all samples considered, there has been no indication of unusually small values of $H(q)$ that would conflict with the standard predictions for the hydrodynamic function made by the δγ theory and the ASD simulations.

The paper is organized as follows: Section II summarizes the theoretical background and explains our OMF model calculations. Experimental details on the silica in DMF system, and the experimental methods and setup are provided in section III. Section IV includes the discussion of our experimental results and the detailed comparison with theory. Finally, our conclusions are contained in section V.

## II. THEORY

### 2.1 Short-time dynamics

In photon correlation spectroscopy and neutron spin echo experiments on colloidal spheres, the dynamic structure factor,

$$S(q,t) = \left\langle \frac{1}{N} \sum_{l,j=1}^{N} \exp\{i\mathbf{q} \cdot [\mathbf{r}_l(0) - \mathbf{r}_j(t)]\} \right\rangle \quad (1)$$

is probed. Here, $N$ is the number of spheres in the scattering volume, $r_j(t)$ is the vector pointing to the center of the $j$-th colloidal sphere at time $t$, $q$ is the scattering wave vector, and $<...>$ denotes an equilibrium ensemble average. At short correlation times where $\tau_B \ll t \ll \tau_I$, $S(q,t)$ decays exponentially according to[2, 3]

$$\frac{S(q,t)}{S(q)} \approx \exp[-q^2 D(q)t], \quad (2)$$

with a wavenumber-dependent short-time diffusion function $D(q)$. Application of the generalized Smoluchowski equation describing the overdamped colloid dynamics leads to the well-known expression[3]

$$D(q) = D_0 \frac{H(q)}{S(q)}, \qquad (3)$$

which relates $D(q)$ to the hydrodynamic function, $H(q)$, and to the static structure factor $S(q) = S(q, t = 0)$. Thus, the hydrodynamic function can be determined from combing a short-time XPCS experiment for $D(q)$ with a SAXS measurement of $S(q)$. The microscopic expression for $H(q)$, obtained from the generalized Smoluchowski equation, is given by[27]

$$H(q) = \left\langle \frac{k_B T}{N D_0} \sum_{l,j=1}^{N} \hat{\mathbf{q}} \cdot \mu(\mathbf{r}^N)_{lj} \cdot \hat{\mathbf{q}} \exp[i\mathbf{q} \cdot (\mathbf{r}_l - \mathbf{r}_j)] \right\rangle, \qquad (4)$$

where $\hat{\mathbf{q}}$ is the unit vector along the direction of $\mathbf{q}$, and the $\mu(\mathbf{r}^N)_{lj}$ are the hydrodynamic mobility tensors relating a force on sphere $j$ to the resulting velocity of sphere $l$. The mobility tensors depend in general on the positions, $\mathbf{r}^N$, of all $N$ particles. This makes an analytic calculation of $H(q)$ intractable, so that approximations have to be introduced.

The positive-valued function $H(q)$ contains the influence of the HI on the short-time diffusion. It can be expressed as the sum of a $q$-independent self-part, and a $q$-dependent distinct-part according to

$$H(q) = \frac{D_s}{D_0} + H_d(q), \qquad (5)$$

where $D_s$ is the short-time translational self-diffusion coefficient.

In the large-$q$ limit, the distinct part vanishes and $H(q)$ becomes equal to $D_s/D_0$. Without HI, $H(q)$ is equal to one for all values of $q$. Any variation in its dependence on the scattering wave number is thus a hallmark of the influence of HI. The function $H(q)$ has the physical meaning of a (reduced) short-time generalized mean sedimentation velocity in a homogeneous suspension that is subject to weak force field, collinear with $\mathbf{q}$ and oscillating spatially like

cos(**q**·**r**). Therefore, $H(q \rightarrow 0) = U_s/U_0$ is equal to the concentration-dependent short-time sedimentation velocity, $U_s(\phi)$, of a slowly settling homogeneous suspension of spheres taken relative to the sedimentation velocity, $U_0$, at infinite dilution. Consequently, the long-wavelength limit of $H(q)$ can be determined alternatively by means of a macroscopic sedimentation experiment[28, 29].

For values $qa \ll 1$, $D(q)$ reduces to the short-time collective or gradient diffusion coefficient $D_c$. For non-zero concentration and low salt content, $D_c$ can be substantially larger than $D_0$, since the relaxation of long-wavelength density fluctuations is speeded up by the low osmotic compressibility[30, 31].

## 2.2 Non-hydrodynamic interactions

The analytic calculations of $H(q)$ and $S(q)$ discussed in this work are based on the one-component macroion fluid (OMF) model. In this model, the colloidal particles are described as uniformly charged hard spheres interacting by an effective pair potential of Derjaguin-Landau-Verwey-Overbeek (DLVO) type[32]

$$\frac{u(r)}{k_B T} = L_B Z^2 \left( \frac{e^{\kappa a}}{1 + \kappa a} \right)^2 \frac{e^{-\kappa r}}{r}, \quad r > 2a \ . \tag{6}$$

For the closed systems studied here, the electrostatic screening parameter, $\kappa$, is given by

$$\kappa^2 = \frac{4\pi L_B \left[ n|Z| + 2n_s \right]}{1 - \phi} = \kappa_{ci}^2 + \kappa_s^2 \ , \tag{7}$$

where $n$ is the colloid number density, $n_s$ is the number density of added 1-1 electrolyte, and $\phi = (4\pi/3) n a^3$ is the colloid volume fraction of spheres with radius $a$. Furthermore, $Z$ is the (effective) charge on a colloid sphere in units of the elementary charge $e$, and $L_B = e^2/(\varepsilon k_B T)$ is the Bjerrum length for a suspending fluid of dielectric constant $\varepsilon$ at temperature $T$. We note that $\kappa^2$ is the sum of a contribution, $\kappa_{ci}^2$, due to surface-released counter-ions, which are monovalent for silica spheres in DMF, and a contribution, $\kappa_s^2$, arising from the added 1-1 electrolyte which is LiCl in our case. The factor $1/(1-\phi)$ corrects for the free volume accessible to the microions, owing to the presence of colloidal spheres[33, 34]. It is of relevance for dense suspensions only.

For strongly charged spheres where $L_B |Z|/a >1$, the value of Z in Eq. (6) for the pair potential needs to be interpreted as an effective or renormalized charge number smaller than the bare one, that accounts to some extent for the non-linear electrostatic screening close to the colloid surface caused by quasi-condensed counterions. Likewise, a renormalized value for κ must be used. Several schemes have been developed to relate the effective Z and κ to the bare ones[35, 36, 37, 38, 39, 40, 41]. The outcome of these schemes depends to some extent on the approximation made for the (grand)-free energy functional, and on additional simplifying model assumptions such as the usage of a spherical cell[42, 43] or the spherical jellium model description[40].

There is an ongoing discussion on how the effective charge and screening parameters in the effective pair potential are related to their bare counterparts, and under what conditions three-body and higher-order corrections to the OMF pair potential come into play. Even though promising advances have been made, this remains as a challenging many-particle problem, in particular for low-salinity systems with strongly overlapping electric double layers. Since our major concern is about short-time dynamic properties, we use the OMF model as a well-defined and well-established model that certainly captures essential features of charge-stabilized suspensions, allowing thus to study general trends in the short-time dynamic properties. However, in subsection 4.3, we shall discuss on a qualitative level the salt- and colloid concentration dependence of the effective charge Z, obtained from a static structure factor fit procedure explained below, and compare it with our findings for the electrophoretic colloid charge, and the renormalized colloid charge obtained by a renormalized jellium model calculation.

Analytic calculations of short-time properties based on the δγ scheme require the colloidal static structure factor, $S(q)$, as the only input, but no additional information on the form and parameters of the underlying pair potential. To calculate $S(q)$ on the basis of the OMF pair potential in Eq. (6), we use the well-established Rogers-Young (RY) and rescaled mean spherical approximation (RMSA) integral equation schemes[3, 20, 21, 44]. The RY scheme is known for its excellent structure factor predictions for fluid systems of Yukawa-like particles. The RMSA results for $S(q)$ are in most cases nearly identical to the RY predictions, provided a somewhat different and usually larger value of $Z$ is used. RMSA calculations, in particular, are very fast, and can be efficiently used when extensive structure factor scans are required over a broad range of parameters.

**2.3 Method of calculation**

The most comprehensive analytic method available to date that allows to predict the hydrodynamic function of dense suspensions of charge-stabilized or neutral hard spheres, is the (zeroth order) δγ-scheme of Beenakker and Mazur[12]. It is based on a partial resummation of the many-body HI contributions. According to this scheme, $H(q)$ is obtained to leading order in the renormalized density fluctuations from[12, 15]

$$H_d(q) = \frac{3}{2\pi} \int_0^\infty d(ak) \left(\frac{\sin(ak)}{ak}\right)^2 [1+\phi S_{\gamma 0}(ak)]^{-1} \times \int_{-1}^{1} dx\, (1-x^2)[S(|q-k|)-1] \quad , \tag{8}$$

and

$$\frac{D_s(\phi)}{D_0} = \frac{2}{\pi} \int_0^\infty dx \left(\frac{\sin x}{x}\right)^2 [1+\phi S_{\gamma 0}(x)]^{-1} , \tag{9}$$

where $x$ is the cosine of the angle extended by the wave vectors $\mathbf{q}$ and $\mathbf{k}$, and $S_{\gamma 0}(x)$ is a known function independent of the particle correlations and given in Refs.[12, 15]. The only input required is the static structure factor $S(q)$, which we calculate using the RY and RMSA integral equation schemes. As can be noted from Eq. (9), $S(q)$ enters only into the distinct part of $H(q)$ since, to lowest order in the renormalized density fluctuations expansion, the self-part is independent of $S(q)$. For charged spheres, the short-time self-diffusion coefficient is thus more roughly approximated by the value for neutral hard spheres at the same volume concentration $\phi$, independent of the sphere charge and screening parameter. To include the actual pair correlations into the calculation of $D_s$ requires to go one step further in the fluctuating density expansion, which severely complicates the scheme. Yet, a detailed comparison of the $\delta\gamma$-scheme predictions with ASD simulations and experimental data on charge-stabilized systems has shown that $H_d(q)$ is in general well captured by Eq. (8). This observation allows to improve the $\delta\gamma$ scheme through replacing the $\delta\gamma$- $D_s$, say, by a more accurate simulation prediction, which is computationally less expensive than a full simulation of $H(q)$. The $\delta\gamma$-$H(q)$ is then shifted upwards by a small to moderately large amount, depending on the system salinity, owing to the fact that the $D_s$ of charged spheres is larger than for neutral ones[3, 25, 45]. In deionized suspensions of strongly charged spheres, $D_s$ scales with the concentration according to $D_s/D_0 = 1 - a_t\,\phi^{4/3}$, with an amplitude factor $a_t \approx 2.5$[ref. 25]. However, even without an improved input for $D_s$, the $\delta\gamma$ scheme remains useful in predicting general trends in the behavior of $H(q)$ on a semi-quantitative level of accuracy. The analytic simplicity of the $\delta\gamma$-scheme has allowed us in the present work to interpret theoretically a comprehensive set of short-time diffusion data for a large range of system parameters.

## III. MATERIALS AND EXPERIMENTAL METHODS

### 3.1 Sample characterization

In the present series of experiments, we use uncoated silica particles dispersed in dimethyl formamide (DMF). Monodisperse silica seed particles with a hydrodynamic radius of 27 nm were synthesized in a microemulsion[46, 47]. These seed particles were grown to their final size by seeded growth with continuous addition of monomer according to Giesche[48]. The final sample was redispersed in DMF. The solvent DMF was purchased from Sigma (CHROMASOLV Plus, for HPLC, ≥99.9%). A refractive index of $n$ = 1.428, mass density of $\rho_{DMF}$ = 0.944 g/cm$^3$, shear viscosity $\eta_0$(DMF) = 0.92 cP and dielectric constant $\varepsilon_{DMF}$ = 36.7 at 20°C were used for DMF to calculate the hydrodynamic radius (PCS), the colloid volume fraction and the Debye screening length. For the silica particles, a mass density of $\rho_s$ = 1.95 g/cm$^3$ was assumed for further calculations. The batch suspension of silica particles in DMF was prepared by performing multiple sequences including the centrifugation of a dilute sample at 1000 g for 10 hours, the exchange of the supernatant with pure DMF and a subsequent redispersion. In the final step of the preparation, the supernatant was removed to such a level that the colloid weight concentration of the suspension was 0.35 g/g. The silica particles were then resuspended by intensive mechanical mixing.

Transmission electron microscopy gave an average particle radius of 80.5 nm, with a relative standard deviation of the size distribution smaller than 2%. For comparison, the hydrodynamic radius of the spheres as measured by dynamic light scattering in a very dilute DMF suspension amounts to 87.7 nm. The radius determined from the form factor measurement in dilute DMF suspensions using SAXS and a homogeneous-spheres form factor fit is 85.5 nm, again with a size polydispersity smaller than 2 %. The differences between these various particle radii are in fact rather small in comparison to the differences found in some other silica dispersions[49].

To control the ionic strength, we added controlled amounts of a lithium chloride (LiCl, Sigma, ≥ 99 %) solution in DMF (200 mM batch solution).

In our X-ray scattering experiments, the samples were filled into quartz capillaries of a diameter of 1.5 mm (0.01 mm wall thickness) and sealed with glue.

**3.2 Experimental methods**

The X-ray scattering experiment was performed at the Troïka III part of the ID10A beamline of the European Synchrotron Radiation Facility (ESRF) in Grenoble. The experimental details concerning the beam characteristics and the general setup have been given elsewhere[24, 50], and can be also found on the ESRF web page (http://www.esrf.eu). We took advantage of the uniform filling mode of the synchrotron which allowed us to avoid the problem of oscillations in the correlation functions[50]. The average current of the beam was about 200 mA. For all the measurements, we utilized radiation of the wavelength of 1.55 Å, which corresponds to a photon energy of 7.99 keV.

The size of the pinhole placed 25 cm in front of the sample was a compromise between the need for high flux ($10^9$ photons/s/100 mA) and the beam coherence. For the XPCS measurements, it was reduced to 12 μm, while for the SAXS scans it could be opened to twice this value. The capillary containing the sample was mounted in a chamber connected to a long pipe guiding the scattered photons to the detector placed at the distance of 3.480 m from the sample. Both the chamber and the pipe were evacuated. The two slits (vertical and horizontal) in front of the detector (Bicron scintillation counter) were closed to 150 μm to provide a reasonable coherent and unsmeared detection of scattered radiation in the XPCS measurements. For the SAXS scans the vertical size of the detector window was usually increased to about 300 μm. Time correlation functions were calculated using an ALV-6000 digital correlator (ALV GmbH, Langen).

Electrophoresis measurements were performed on a Malvern Zetasizer 2000. With this apparatus, we have measured the electrophoretic mobility of silica dispersions in DMF at 25 °C as a function of the LiCl concentration, and for a particle concentration of 2.5 g/l, corresponding to a volume fraction of about $1.25 \times 10^{-3}$. The measurements were repeated several times to check for the reproducibility of the results.

## IV. RESULTS AND DISCUSSION

In the following, we discuss our experimental results for the static and short-time diffusion properties of fluid-like suspensions of silica spheres in DMF (at fixed temperature T = 20 $^{o}$C). We investigated three distinct series of samples labeled by H, K and C, respectively. All parameters characterizing these series are summarized in the Table 1. In the first two series H and K listed in the table, the weight concentration, $C_w$, quantifying the weight of silica spheres relative to that of the whole suspension, was fixed to 150 mg/g and 250 mg/g, respectively, while the concentration, $C_s$, of added LiCl was varied. In the less concentrated samples of series H, $C_s$ was varied from 0 – 20 mM, while $C_s$ was varied from 70 – 400 µM in the more concentrated samples of series K. Series C consists of samples with fixed salinity $C_s$ = 50 µM, with the weight concentration of silica spheres varied from 100 – 200 mg/g. For all samples listed in the table, we have inspected by eye that there is no indication of crystallization in form of iridescent light. In series H, some traces of crystallization were recognized by the appearance of crystal-like peaks in the SAXS scattering function within the added salt region of 5 – 10 µM. In series K, a reduction of the salt concentration below 70 µM resulted in the crystallization of the samples.

In the present paper, systems indicative of (partial) crystallization have been excluded from the analysis, and from Table 1, since our focus here is on liquid-like samples. The XPCS analysis of the dynamics in (partially) crystalline samples requires a special analysis, different from that for liquid-like systems. This will be discussed in a forthcoming paper.

The effective charge numbers, $Z_{RY}$ and $Z_{RMSA}$, given in the table have been deduced by fitting the peak height of the RY and RMSA $S(q)$, respectively, to the experimental one. The experimental structure factors (in Figs. 2 – 4) were first fitted using the very efficient RMSA code. In a second step, the more elaborate and less time-efficient RY code was applied to fit the structure factor peaks by adjusting $Z_{RY}$, with $Z_{RMSA}$ providing an upper limit to the actual charge number. It is well-known that the RMSA values for the effective charge are in most cases somewhat larger than the more accurate RY values, reflecting the tendency of the RMSA to underestimate static pair correlations. A small readjustment of the volume fraction, $\phi$, of silica spheres in the samples of series H and K was required for a nearly perfect match of the wavenumber position of the RY and RMSA structure factor peaks to the experimental one. This is the reason for the small variation in the volume fraction values of series H and K in the table. Aside from this slight readjustment, the effective charge has been the only adjustable parameter. The values for the screening parameter, $\kappa$, in the table were obtained using Eq. (7) in combination with the RY value of the effective charge.

**Table 1:** Parameters used in the $\delta\gamma$-scheme calculations of $H(q)$ and $D(q)$. The solvent is DMF at $T = 20°C$ and $\varepsilon = 36.7$, corresponding to a Bjerrum length $L_B = 1.554$ nm. The radius a = 85.5 nm as obtained from the form factor measurement was used in all our calculations. The listed values for the effective charge, $Z$, were obtained from fitting the peak values of the RY and RMSA-$S(q)$, respectively, to the experimental ones (see Figs. 2 – 4). Three different series of samples have been studied. The samples in series H and K are for a fixed particle concentration but varying salinity, whereas in series C the salinity is fixed and the silica concentration varied. $\phi$ - silica volume fraction; $C_w$ - silica weight concentration in mg/g; $C_s$ – added LiCl concentration; $Z_{RMSA}$, and $Z_{RY}$ – effective charge numbers, in units of elementary charge $e$, obtained respectively from the RMSA and RY structure factor peak height fits; $\kappa$ - inverse screening length calculated using $Z_{RY}$. $Z_{el}$ – effective charge number deduced from electrophoresis measurements on dilute samples ($\phi = 1.25\times10^{-3}$). The missing input in the table for $C_s = 20$ mM is due to the fact that for such a high amount of salt the peak height of $S(q)$ is quite insensitive to the value of Z. For a discussion of the salinity values between brackets in the K series, see subsection 4.3.

| Sample | $\phi$ | $C_w$ [mg/ml] | $Z_{RMSA}$ [e] | $Z_{R-Y}$ [e] | $Z_{el}$ [e] | $\kappa$ [nm$^{-1}$] | $\kappa a$ |
|---|---|---|---|---|---|---|---|
| $C_s$ [µM] | | | **Series H** $C_w$ = 150 mg/g | | | | |
| 0 | 0.0795 | 155 | 220 | 177 | 383 | 0.0108 | 0.923 |
| 20 | 0.0744 | 145 | 350 | 282 | 845 | 0.0260 | 2.22 |
| 50 | 0.0718 | 140 | 600 | 520 | 895 | 0.0396 | 3.38 |
| 100 | 0.0769 | 150 | 1000 | 900 | 928 | 0.0557 | 4.76 |
| 200 | 0.0820 | 160 | 5000 | 3500 | 1036 | 0.0864 | 7.39 |
| 20 000 | 0.0769 | 150 | -- | -- | 2552 | > 0.686 | > 58.6 |
| $C_s$ [µM] | | | **Series K** $C_w$ = 250 mg/g | | | | |
| 70 | 0.131 | 255 | 1600 | 850 | -- | 0.0534 | 4.56 |
| 100 (80) | 0.131 | 255 | 2500 | 1700 | 928 | 0.0679 | 5.80 |
| 150 (130) | 0.131 | 255 | 3000 | 2200 | -- | 0.0808 | 6.91 |
| 200 (160) | 0.136 | 265 | 4000 | 4000 | 1036 | 0.101 | 8.64 |
| 400 | 0.136 | 265 | 5000 | 3500 | -- | 0.122 | 10.4 |
| $C_w$ [mg/g] | | | **Series C** $C_s$ = 50 µM | | | | |
| 100 | 0.0564 | 110 | 900 | 790 | 895 | 0.0400 | 3.42 |
| 150 | 0.0718 | 140 | 600 | 520 | 895 | 0.0396 | 3.38 |
| 200 | 0.108 | 210 | 380 | 320 | 895 | 0.0401 | 3.43 |

As we have found by visual inspection, there is a broad range of colloid and salt ion concentrations where the silica samples crystallize at least partially, especially at high volume fractions. On the other hand, for too low colloid concentrations, the granted beam time of a few days would have allowed us to probe only a few wave numbers, and even then with unsatisfactory statistics only. Therefore, and since we wanted to study fluid-like samples, we carefully selected two rather high volume fractions of silica spheres, namely those of series H and K given, respectively, by $\phi \approx 0.07 - 0.08$ and $\phi \approx 0.13 - 0.14$. This has allowed us to scan a broad range of added salt concentrations while having samples in the fluid state. In this way, we could tune the electrostatic interactions from being practically screened out to very strong values, for a single fixed colloid concentration.

**4.1 Form factor**

Fig. 1 shows the results of our SAXS measurements of the mean scattered intensity, $I(q)$, obtained in a broad range of wavenumbers $q$, for a silica-sphere suspension of moderate concentration of 150 mg/ml, and in presence of 100 μM added LiCl. Under the conditions of this experiment, the structure factor levels off at the value of one, within the experimental errors, already for $q > 0.08$, which defines thus the region where $I(q) \cong P(q)$. The experimental points were scaled to match the theoretical form factor curve, $P(q)$, for a Gaussian diameter distribution of homogeneous spheres of mean radius 85.5 nm, and relative standard deviation of 1 %. The presence of multiple and sharp minima in the experimental data is a clear indication of a very low size polydispersity, quantified by the small value of 1 % for the relative standard deviation in the calculated $P(q)$.

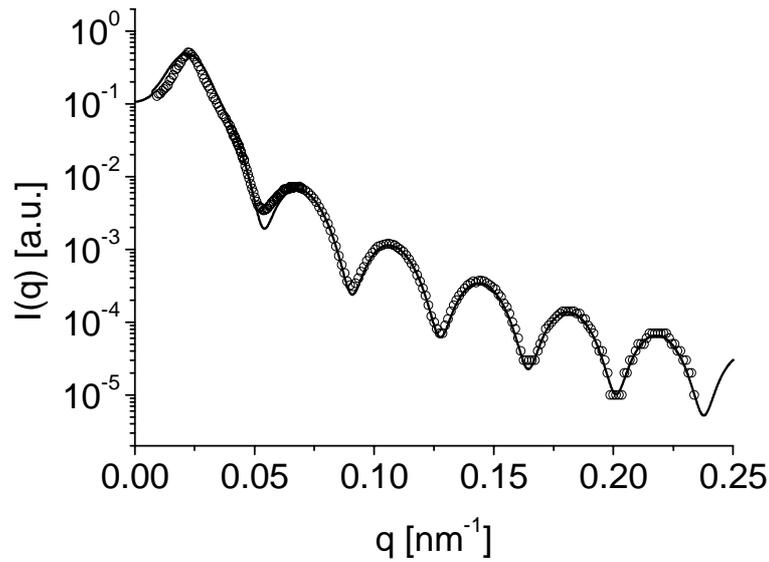

FIG. 1: SAXS result for the normalized mean scattered intensity, obtained for silica spheres in DMF at the concentration of 150 mg/ml and with 100 μM added LiCl (open circles). The solid line represents the calculated intensity $I(q) = S(q) P(q)$, where the form factor, $P(q)$, was calculated for a Gaussian size distribution of homogeneously scattering spheres of mean radius 85.5 nm and relative standard deviation of 1 %.

**4.2 Structure factor and short-time diffusion properties**

For each sample, a static SAXS measurement was performed for an extended number of scattering wavenumbers in order to identify the position and height of the static structure factor peak. The analysis of the SAXS data to deduce $S(q)$ from the measured mean intensity included a correction for the $q$-dependent background obtained for pure DMF. After this background correction, the system parameters were determined from matching the shape of the product function of the RY or RMSA-$S(q)$ and $P(q)$, to the background-corrected experimental intensity data for the full range of experimental $q$ values.

In the XPCS measurements of the scattered radiation intensity autocorrelation function, $g^{(2)}(q,t)$, six to twelve $q$-values out of a set of wavenumbers probed in the SAXS experiments were selected. Then, using the Siegert relation, the first-order correlation function $g^{(1)}(q,t)$ was calculated and analyzed in terms of the (short-time) initial slope, $\Gamma_i(q)$, of $-\ln g^{(1)}(q,t)$. The values for the short-time diffusion function were obtained from $D(q) = \Gamma_i(q) / q^2$, and then normalized by the value $D_0 = 2.72\times10^{-8}$ cm$^2$/s for the free diffusion coefficient of a single silica sphere obtained from a light scattering PCS experiment.

In Figs. 2 – 4, we have plotted the reciprocal of the normalized diffusion function, $D_0/D(q)$, rather than $D(q)$ itself. This allows for a direct comparison with the data for $S(q)$ since $D_0/D(q)$ reduces to the static structure factor for negligible HI. Thus any observed difference in the two functions indicates the influence of HI. The experimental data points for $H(q)$ were obtained from multiplying the experimental $D(q)/D_0$ by the peak-height fitted RY-$S(q)$ instead of the experimental points to avoid progression of statistical errors in the data. This procedure is justified since for all samples considered in Figs. 2 - 4, the RY-S(q) is a decent description of the experimental structure factors. The $\delta\gamma$-scheme results for $H(q)$ shown in the figures have been calculated using the peak-height- fitted RY $S(q)$, which is the only input in the Eqs. (8-9).

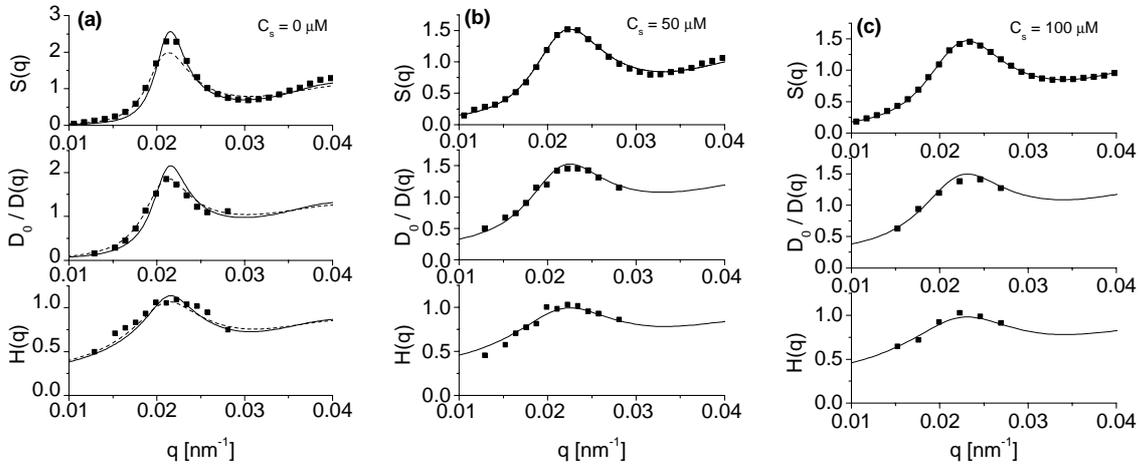

FIG. 2 (a-c): Results for $S(q)$, $H(q)$ and the reciprocal of the normalized $D(q)$, obtained for the fixed colloid concentration samples of series H. The filled symbols are the data from the combined SAXS and XPCS measurements. The lines are RY-fit results of $S(q)$, and the $\delta\gamma$-scheme results for $H(q)$ and $D(q)$ based on the RY-$S(q)$ input. The dashed lines for $S(q)$ and $H(q)$ in Fig. 2a are the RY and $\delta\gamma$-scheme results obtained when the peak height of the experimental diffusion function $D(q)$ is fitted instead of the peak in $S(q)$ (see the text for details). All system parameters used in the calculations are listed in Table 1. Additional results on the series H are available in the supplementary material in Fig. S1.

Fig. 2 includes a summary of results for the samples of series H obtained from the SAXS measurements of $S(q)$, and the XPCS measurements of $D(q)$. The concentrations of added salt are indicated in the upper right corner of the plots. Additionally shown is the hydrodynamic function, $H(q)$, as obtained from the experimental data for $D(q)$ by using Eq. (3) (symbols), and from $\delta\gamma$-scheme calculations by using the RY input for $S(q)$ (lines). In the low-salt region interval of series H in between 5 – 10 µM, which we have explored also experimentally, we find some traces of crystallization. In particular, for $C_s$ = 5 µM, the measured scattered intensity contained crystal-like peaks and could thus not be well fitted using the RY and RMSA integral schemes, which are designed to describe the fluid state only. Therefore these data are not shown. While the zero-salinity sample ($C_s$ = 0 µM) in series H appears to be fluid-like and has therefore been included into Table 1, it can not be excluded that it is partially crystalline. In fact, if instead of $S(q)$ one alternatively fits the peak height of the $\delta\gamma$-RY $D(q)$ to the experimental diffusion function, the resulting RY-$S(q)$ turns out to be less structured than the experimental one (see the dashed lines in Fig. 2a), indicating that there is probably some additional intensity contribution to the structure factor peak that arises from partially crystalline order.

Due to the dissociation-association mass action equilibrium of the silanol (SiOH) groups on the silica surfaces, the number of surface charges (i.e., SiO⁻ groups) can decrease with decreasing salt concentration[1, 51]. Nevertheless, the effect of the larger electrostatic screening length dominates for very small values of $C_s$ causing (partial) crystallization. This interesting point will be further discussed in a forthcoming article where we shall study the structure and dynamics of (partially) crystalline colloidal systems.

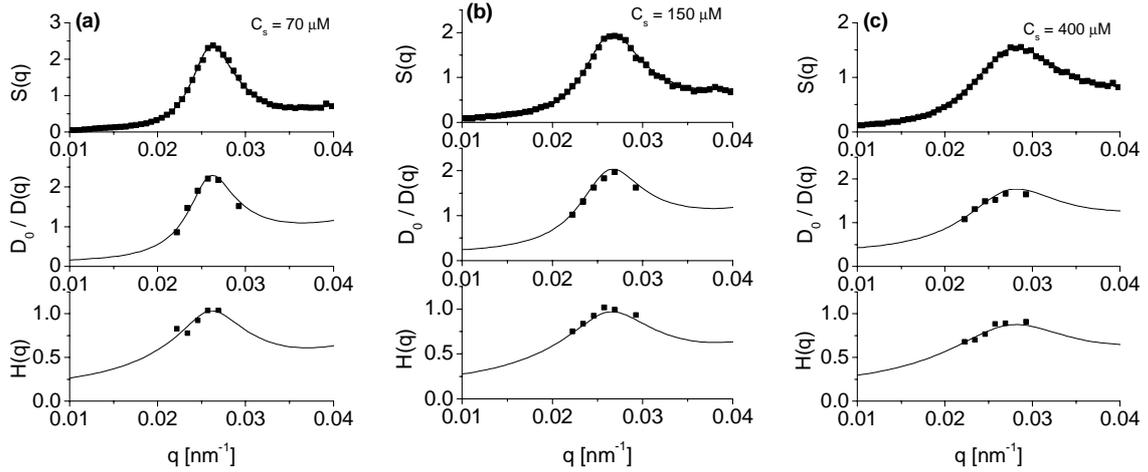

FIG. 3 (a-c): As in Fig. 2, but for the more concentrated samples of series K. The salt concentrations, $C_s$, are indicated in the plots. Additional results on this series are available in the supplementary material in Fig. S2.

A corresponding set of results for the more concentrated samples of series K, where $\phi \approx 0.13 - 0.14$, is included in Fig. 3. Moreover, the results for the samples of series C of fixed salinity $C_s = 50$ µM, and colloid concentration varying from 100 to 200 mg/g, are depicted in Fig. 4.

The samples in series C could be maintained in the fluid state only for this comparatively large amount of added salt. As can be noticed from Figs. 2 – 4, the undulations in $H(q)$ are growing with increasing $\phi$ and decreasing salt content, reflecting a similar behavior in $S(q)$. The peak positions of the two functions $S(q)$ and $H(q)$, and thus of $D_0 / D(q)$, are all located at the same wavenumber $q_m$, in accord with the theoretical prediction. The diffusion function, $D(q)$, attains its minimal value smaller than $D_0$ also at $q_m$. This value of the so-called cage-diffusion coefficient, $D(q_m)$, describes the most slowly progressing (initial) decay of density fluctuations at a wavelength, $2\pi/q_m$, quantifying the spatial extent of the dynamic cage of next-neighbor particles. With increasing particle correlations, e.g., increasing $\phi$ or decreasing $C_s$, the dynamic cage stiffens and $D(q_m)$ becomes smaller. This is reflected in a rising peak value of the inverse of $D(q_m)$ with growing particle correlations.

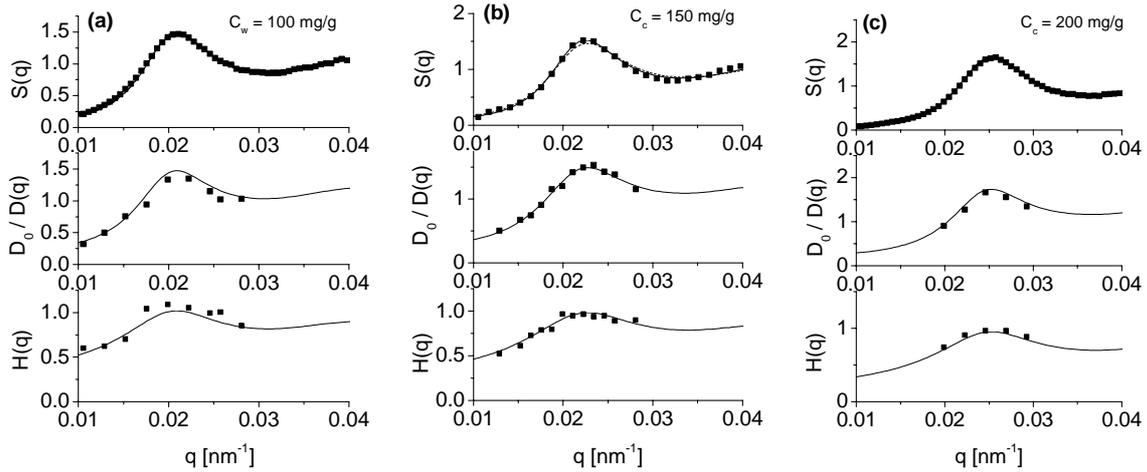

FIG. 4 (a-c): $S(q)$, $H(q)$ and the reciprocal of the normalized $D(q)$, for the fixed salt concentration samples of series C, with $C_s = 50$ μM maintained fixed. The filled symbols represent the combined SAXS and XPCS data, and the lines are the RY-fit results for $S(q)$, and the δγ-theory results for $H(q)$ and $D(q)$. The values of the silica weight concentration, $C_w$, are indicated in the plots. Parameters used in the calculations are given in Table 1.

The consistently good agreement between the experimental data for $D(q)$ and the δγ-scheme results, which is observed for practically all fluid-like samples in this study, is quite remarkable, considering that only the experimental structure factor peak height has been fitted. The agreement remains good even for the low-salinity samples, unless these exhibit significant traces of crystalline order.

The low-salt systems in our study are precisely in the range of volume fractions where, for some charge-stabilized colloidal systems, extraordinarily small values of $H(q)$ have been reported[8, 9, 10, 11], which conflict with the predictions of the δγ-theory and Stokesian Dynamics simulations[23, 24, 25]. In the present study, and in many others[18, 19, 22, 23, 24, 52, 53], such conflicting experimental results for $H(q)$ do not occur.

### 4.3 Salt-concentration dependence of various quantities

In the special case of series K, we have made an interesting observation. On trying to fit the SAXS scattered intensity in this series using the RMSA scheme, we noticed that the peak height of the RMSA intensity could not be enlarged enough to reach the experimental intensity peak values, regardless of the selected value of $Z_{RMSA}$. For a certain $Z_{RMSA}$, a maximal peak height of the RMSA- $S(q_m)$ is reached, from which on $S(q_m)$ becomes smaller again when $Z_{RMSA}$ is further increased. The maximal peak height in S(q) as a function of Z is reached when the electrostatic screening effect by the surface-released counterions which becomes stronger with increasing *Z,* starts to over-compensate the effect of the rising contact value of the OMF potential. The experimental structure factor peak value could be reached using the RMSA only by additionally lowering the concentration of added salt to values below the experimental one (these are the $C_s$ values between brackets listed in Table 1). However, since the maximum in $S(q_m)$ as a function of the effective charge is larger in the RY approximation than in the RMSA scheme, it was possible to fit the experimental data by the RY scheme without deviating from the experimentally given salt concentration. This finding also highlights the advantage of the more accurate RY scheme over the RMSA, in that certain more strongly coupled systems can be described consistently only by this more elaborate, partially self-consistent scheme.

In Fig. 5a, we show the salt concentration dependence of $Z_{RY}$ for all samples in series H and K. There is a certain ambiguity in the effective charge value for the largest salinity considered, since quite different values of the effective charge *Z* in the OMF pair potential can result in practically the same peak height in $S(q)$. The effective charge in series H and K rises steeply in the vicinity of salt concentration values where the number of added salt ions begins to exceed the number of surface-released counterions. This characterizes the transition regime from counterion-dominated to salt-ion dominated screening, where typically the most significant changes take place of quantities characterizing charge-stabilized suspensions.

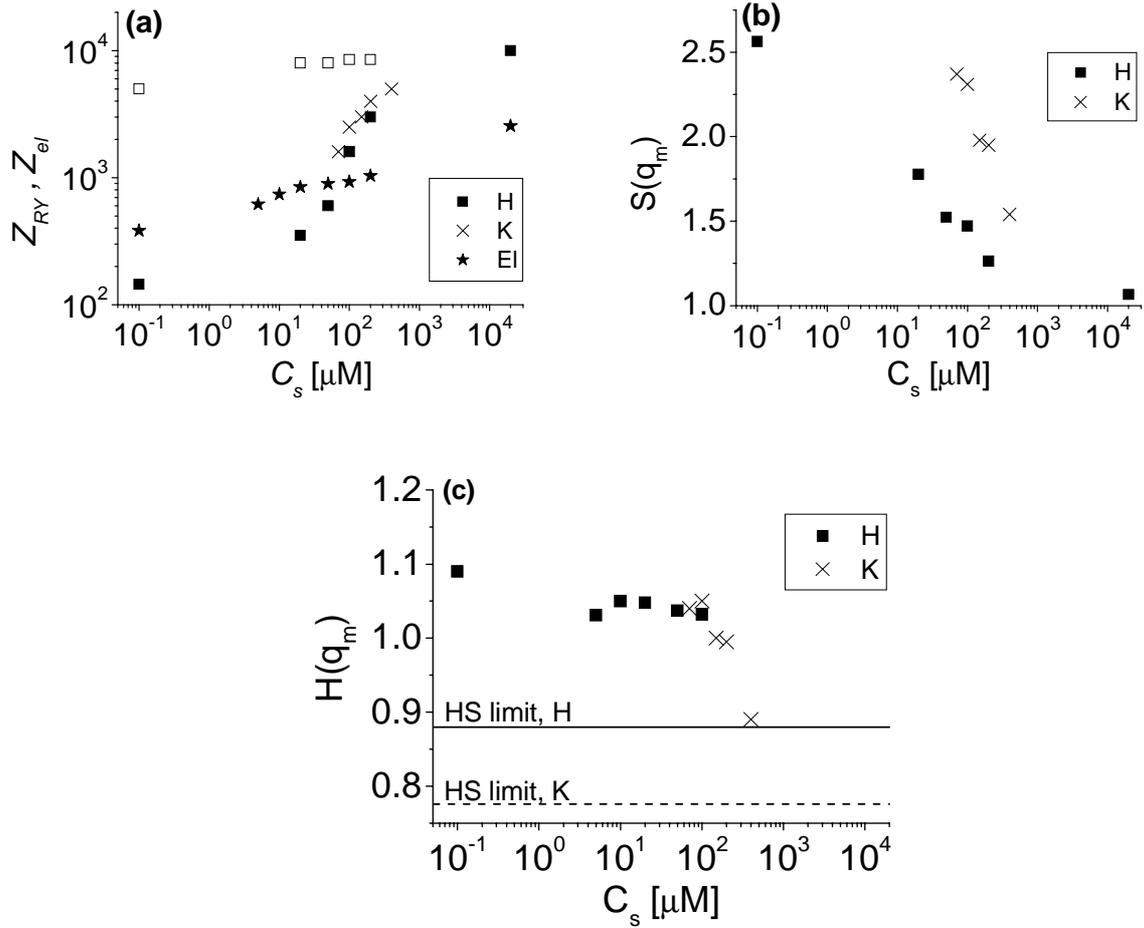

Fig. 5: (a) Fit values of the effective charge number, $Z_{RY}$, and the estimated electrophoretic charge number $Z_{el}$ (★), and (b) maximum of the experimental $S(q)$, and (c) maximum of the δγ-scheme-fitted $H(q)$ of series H (■) and K (✕) as a function of added salt concentration. In (a), also the effective charge fit values from the (unphysical) high-charge branch of series H are included (□). The hard sphere (HS) limits of $H(q_m)$ for series H (solid line) and K (dashed line) are also shown in (c).

At this point of our discussion, it is interesting to note that nearly equally good fits to the static and dynamic scattering data can be achieved using an alternative branch of much higher effective charge values $Z_{RY}$. The charge values of this alternative branch are depicted in Fig. 6 as open squares. The salt-concentration dependence of these effective charges is much weaker since for such high charge values the salt-dominated concentration regime is not reached in the range of $C_s$ values shown in Fig. 5. To find out whether this alternative branch of high-charge values makes sense physically, we have additionally performed electrophoretic light scattering measurements on dilute samples (with $\phi \approx 1.25 \times 10^{-3}$) of the electrophoretic mobility of the silica spheres as a function of $C_s$. From the measured mobility data, a rough estimate of the electrophoretic particle charge number, $Z_{el}$, was obtained using Henry's formula of single-sphere electrophoretic theory[1, 54, 55], by identifying the zeta potential with the electric potential on the sphere surface. The so-obtained values for $Z_{el}$ are displayed as stars in Fig. 5a.

It can be expected, and it was found indeed in recent electrophoresis studies on interacting colloidal particles, that $Z_{el}$ should not be dramatically different from the effective charge appearing in the OMF potential[56, 57]. As one can notice from Fig. 5a, the values of the alternative branch for $Z_{RY}$ are exceedingly larger than the electrophoretic charges, and also larger than the effective charge values of the 'regular' branch listed in the table. The high-charge branch in Fig. 5a is additionally recognized as unphysical by comparing its values with the effective colloid charge numbers estimated using a single-colloid charge-renormalization scheme. For this purpose, we have employed the renormalized jellium model (RJM) scheme[40], which provides values for the renormalized charge, $Z_{ren}$, as a function of the bare one. In comparison with the standard Poisson-Boltzmann cell model scheme for the effective charge by Alexander et al.[42], the RJM has the advantage of being directly linked to the OMF pair potential in Eq. (6), and thus to $Z_{RY}$, and to maintain the validity of Eq. (7) in that the bare charge needs just to be replaced by the effective one. The effective charges derived from the two charge renormalization schemes are not very different from each other and reveal the same trends in their $C_s$ and $\phi$ dependence[35]. For the present silica in DMF systems, we have found that the RJM renormalized charge is consistently smaller than 2000 e even in the saturation limit. Moreover, the renormalized charge shows a weak rise with increasing $C_s$, similar to the electrophoretic charge in Fig. 5a.

For all these reasons, we can rule out the high-charge set of RY values in Fig. 5a. When performing extended structure factor fits using automatic routines, it is important to make sure not to end up with effective charge values of the unphysical branch. This might happen for certain start values in the fitting routine. However, it can not be ruled out for systems different from the present one that the effective charges of the upper branch might be of physical relevance.

We notice from Fig. 5a that $Z_{el}$ is growing with increasing $C_s$, similar to the $Z_{RY}$ from the physical branch. However, the rise in $Z_{RY}$ is far more steep. An effective charge growing with increasing $C_s$ is also predicted by the charge renormalization schemes, and can be explained by the enlarged salt-ion screening that reduces the fraction of counterions that quasi-condense on the colloid surfaces. While the ascent of the three discussed types of effective charges with increasing $C_s$ is expected on physical grounds, we do not understand to date why $Z_{RY}$ in Fig. 5a is far more steeply rising than the other two charges. To gain more insight into the interrelation of the effective colloid charges, charge regulation[58,59] and finite colloid concentration effects[60,61,62,63] should be considered. These effects have not been accounted for in the present discussion of Fig. 5a. The salt-concentration dependence of $S(q_m)$ and $H(q_m)$ in series H and K, is depicted in Figs. 5b and c, respectively. The peak height in $S(q)$ shows the expected drop with increasing salt content, which reflects the corresponding loss in the particle correlations. Likewise, $H(q_m)$ approaches the corresponding hard sphere value.

### 4.4 Colloid-concentration dependence of $H(q_m)$ and $S(q_m)$

The data for $H(q_m)$ in series H and K that are shown in Fig. 5b as a function of $C_s$, and the data for $H(q_m)$ in series C shown in Fig. 6c as a function of the fitted colloid weight concentration, $C_w^{fit}$, are consistent with the generic ordering relation, $H^{CS}(q_m;\phi) > H^{HS}(q_m;\phi)$, predicted both in the Stokesian Dynamics simulations and the δγ-scheme. This relation expresses that the peak height of the hydrodynamic function of charged spheres (CS) is higher than the one of neutral hard spheres (HS) at the same volume fraction[23,24,25]. The peak height value of hard spheres, which provides a lower boundary for charged colloidal spheres, is given to very good accuracy by the linear relation

$$H(q_m) = 1 - 1.35\phi. \tag{10}$$

This relation is valid up to the freezing transition of hard spheres at $\phi = 0.49$ (Refs [25, 64]). We note that the exceptionally small values of $H(q)$ purportedly obtained in certain experiments on low-salt suspensions strongly violate this ordering relation.

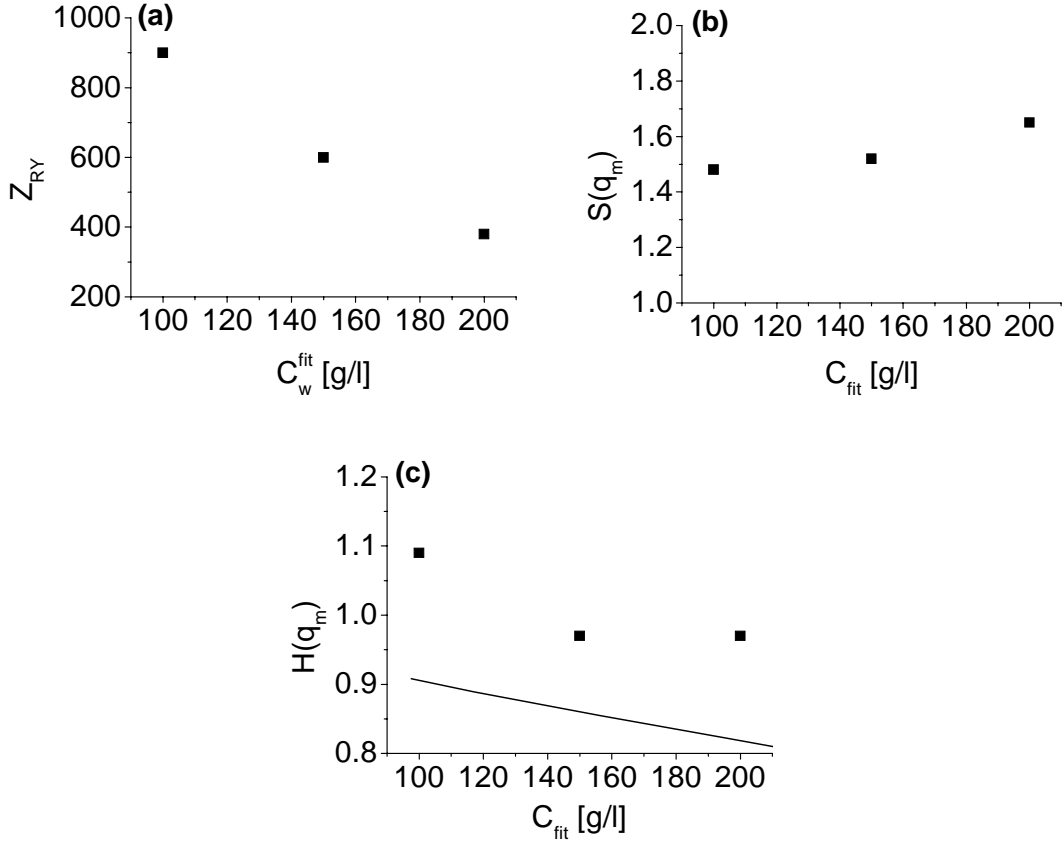

Fig. 6: Series C results (filled symbols) for (a): fit values of the RY effective charge $Z_{RY}$, (b): maximum of $S(q)$ and (c): maximum of $H(q)$ as a function of colloid concentration, for a fixed value of $C_s = 50$ μM, of added salt. The solid line in (c) represents the hard-sphere result, $H(q_m) = 1 - 1.35\phi$.

The effective charge of series C depicted in Fig. 6a declines with increasing weight concentration, i.e., when $\phi$ is increased from 0.056 to 0.108 (see Table 1). This descent in $Z_{RY}$ is compatible with a corresponding descent in the renormalized charge obtained from jellium model calculations for such a system of intermediate salinity[35, 40]. The structure factor peak in series C increases only weakly with growing weight concentration (see Fig. 6b), since the effect of increasing $\phi$ is partially balanced by the decreasing effective charge.

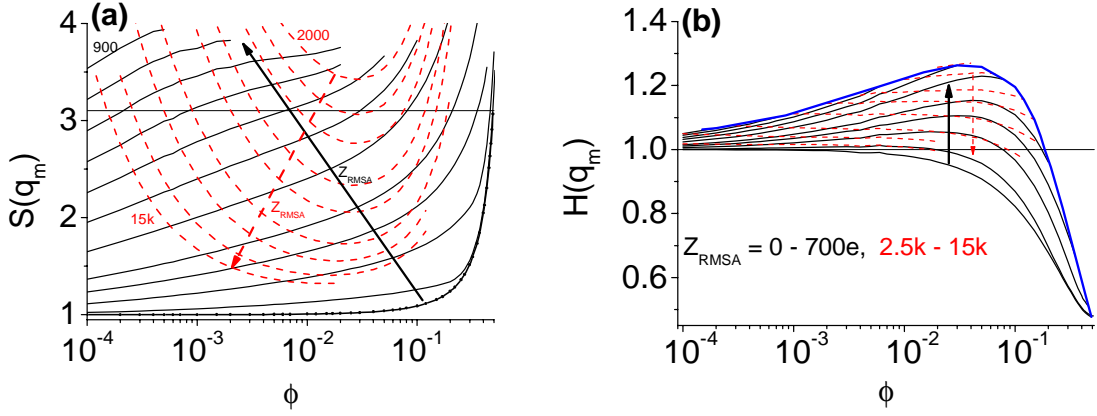

Fig. 7: δγ-RMSA values of (a) $S(q_m)$ and (b) of $H(q_m)$ at salt-free conditions, calculated as a function of $\phi$ for various values of $Z_{RMSA}$ as indicated (iso-$Z_{RMSA}$ lines). The upper limiting contour line in Fig. 7b in blue color is derived from the cutoff condition $S(q_m) \leq 3.1$ for the onset of freezing. The dashed iso-charge lines are obtained from the (unphysical) upper branch of RMSA effective charge values. The solid (dashed) arrows indicate growing values of effective charges from the lower (upper) branch of RMSA fit values.

At constant effective charge $Z$ the highest value of $H(q_m)$ in a given system is attained, according to theory, for a zero amount of added salt. In order to fully explore the general behavior of $H(q_m)$ for fluid suspensions under salt-free conditions ($C_s = 0$), we have used the δγ-scheme with the RMSA static structure factor input, which is numerically efficient enough to do such a detailed exploration in a reasonable amount of time. For each selected value of $Z_{RMSA}$, $H(q)$ was calculated and its maximum value, $H(q_m)$, determined as a function of $\phi$. In parallel, the maxima, $S(q_m)$, of the RMSA-structure factors were calculated (see Fig. 7a) and plotted as a function of $\phi$ for selected effective charges. To limit ourselves to fluid systems, calculated values of $H(q_m)$ have been included in Fig. 7b only when the corresponding value of $S(q_m)$ is not larger than 3.1. This peak value has been reported as a reasonable empirical Hansen-Verlet one-phase criterion[65] for the onset of colloid crystallization[66, 67, 68, 69] of systems with long-range Yukawa-type pair interactions. The value of the peak height where crystallization sets in varies to some extent, depending on the range (softness) of the pair potential, from about 2.85 for hard spheres up to about 3.1 for more dilute samples of Yukawa-type particles with very long-range repulsion (zero-salinity systems). The range of the Yukawa-type pair potential, as quantified by κ*a*, also has a bearing on whether crystallization in a more compact fcc phase or a less compact bcc phase takes place[67, 70, 71].

Accepted values of $H(q_m)$ have been plotted in Fig. 7b in the form of iso-$Z_{RMSA}$ lines. By scanning through increasingly large values of $Z_{RMSA}$, for all of which $H(q_m)$ has been calculated, we have obtained the upper limiting contour line in Fig. 7b that restricts the values of $H(q_m)$ from above for fluid systems. This contour line consists of all iso-charge line points for which $S(q_m) = 3.1$. From below, the set of iso-charge lines of $H(q_m)$ is limited by the curve describing the $H(q_m)$ of neutral hard spheres.

On increasing $Z_{RMSA}$ to a value of about 700, one ends up with the situation where for all volume fractions above $\phi = 10^{-4}$ only (partially) crystalline states, identified by peak values in $S(q)$ larger than 3.1, are encountered. However, we additionally found from our $\delta\gamma$-RMSA analysis that a further increase in $Z_{RMSA}$ brings the system eventually back to the fluid state in the region of moderate volume fractions ($0.001 < \phi < 0.1$). This observation is intimately related to the high-charge values branch of effective charges discussed already in relation to Fig. 5a, that allowed for an alternative fit of the static and short-time experimental data. According to our calculations, the iso-charge lines in Fig. 7b determined for $Z_{RMSA}$ values from the lower (physical – solid lines) and upper (unphysical – dashed lines) branches cover the same region in the $\phi$ - $H(q_m)$ space. In other words: for a given volume fraction $\phi$ each $H(q)$ line obtained using "physical" values of Z has its "unphysical" analog with an identical maximum value $H(q_m)$. Since the addition of salt reduces the degree of order, we can conclude that for any selected pair of values $\phi$ and $C_s$ in the present system, $H(q_m)$ is restricted to values bounded from above by the limiting contour line for zero-salt systems derived from the freezing criterion, and by the zero-charge contour line of neutral hard spheres from below.

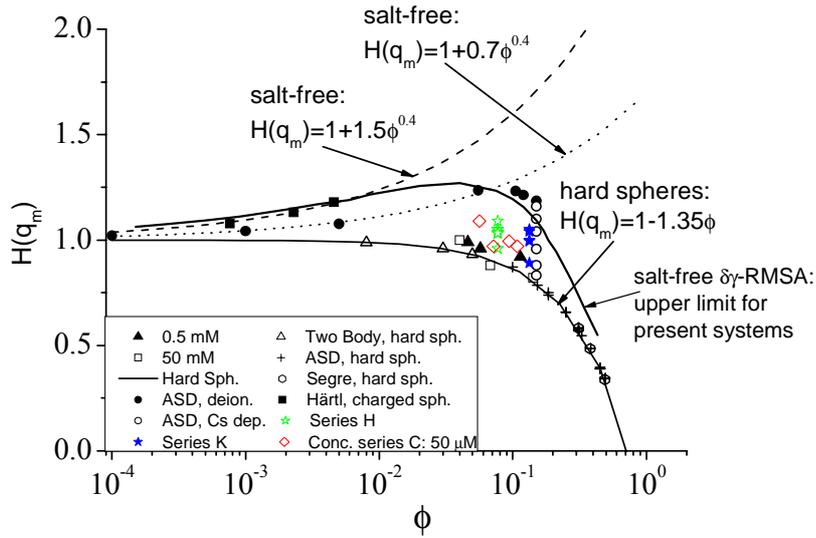

Fig. 8: Peak values of the hydrodynamic function, $H(q)$, as a function of the colloid volume fraction $\phi$. Included are our data for series K, H and C (colored symbols), together with a collection of other experimental and computer simulation data. Lower solid line: hard spheres. Upper solid line: limiting contour line for deionized, fluid systems redrawn from Fig. 7. Dashed line: low-density form of a deionized suspension (with experimental data of Härtl et al.[53], overlayed (■)). Dotted line: low-density form of a deionized system studied in accelerated Stokesian Dynamics (ASD) simulations[24]. Further shown are XPCS data for 0.5 mM (▲) and 50 mM (□) of added 1-1 electrolyte[24], and dynamic light scattering data (○) of Segre et al.[72]. Additionally displayed are also low-density results with full 2-body HI included[64] (△), ASD simulation data[23] without (●) and with (◇) added 1-1 electrolyte, and ASD data[24, 25] for colloidal hard spheres (+).

The maximum peak height attainable in our silica in DMF system, as predicted in the approximate $\delta\gamma$-RMSA scheme, is given by $H(q_m) \approx 1.27$. Interestingly enough, this value can be reached only at a relatively low colloid concentration ($\phi \approx 0.04$) and a moderate effective charge number of $Z_{RMSA} \approx 400$. We can qualitatively rationalize the occurrence of this maximum by the following argumentation: for lower volume fractions, where freezing of particles is achieved by increasing $Z_{RMSA}$, the concentration of hydrodynamically interacting particles is too small to maximize, by spatial ordering, the collective motion of neighboring particles quantified by $H(q_m)$. For volume fractions that are too large, the enhanced crowding of particles causes stronger near-field hydrodynamic interactions to come into play, with the effect of slowing down the cooperative motion of neighboring particles. The near-field hydrodynamic slowing down at higher volume fractions can be so strong that the $H(q_m)$ attains values smaller than one even for low-salt fluid systems.

In Fig. 8, finally, we have plotted the values of $H(q_m)$ obtained in the present experimental study (colored symbols) together with a collection of experimental and computer simulation data by a number of other researchers. Furthermore, we have included in this figure also the upper limiting contour line shown already in Fig. 7b, which defines in $\delta\gamma$-RMSA approximation the upper limit of values attainable by $H(q_m)$ of the present silica in DMF system. It should be noted that all depicted experimental data fall within the region defined by this upper limiting curve and the hard-sphere curve. Also most of the theoretical and computer simulation data are within the bounds of these two curves. Only a few ASD simulation data points in Fig. 8 are slightly above the upper limiting curve. This can be explained by the approximate nature of the $\delta\gamma$-RMSA scheme which, as we have discussed before, tends to underestimate the peak height in comparison to the accurate ASD simulations[22], with the differences increasing with increasing colloid concentration. Note that our data bridge the gap between the hard-sphere limit and the opposite limiting case of deionized systems of strongly charged particles.

For dilute and salt-free, i.e. deionized, suspensions where far-field 2-body HI prevails, numerical calculations have shown that $H(q_m)$ increases initially sub-linearly in the concentration according to[24, 64]

$$H(q_m) = 1 + p_m \phi^{0.4} \ . \qquad (11)$$

This low-density asymptotic form, however, defines not a unique curve since the coefficient, $p_m > 0$, is moderately dependent on the interaction parameters $L_B Z/a$ and $\kappa a$ (see Fig. 8). The coefficient is larger for more strongly structured suspensions characterized by a higher peak in $S(q)$. In contrast, the exponent 0.4 is independent of the system parameters, as long as the effective particle charge is large enough for the physical hard core of the particles to remain masked by the electrostatic repulsion. Within the small-$\phi$ range where the asymptotic form in Eq. (11) applies (i.e., typically for $\phi < 0.01$), its curves are located below the limiting contour line of the corresponding system.

**SUMMARY AND OUTLOOK**

Using combined XPCS and SAXS measurements, the short-time diffusion functions $D(q)$ and $H(q)$, and the static structure factor $S(q)$ of charged silica spheres in DMF have been systematically studied in a broad range of colloid and salt concentrations. In fact, in contrast to earlier work the explored systems cover the full fluid range from hard-sphere-like systems to the opposite extreme of deionized suspensions. The data for $D(q)$ and $H(q)$ obtained by these measurements for fluid-ordered systems could be consistently described by the $\delta\gamma$-scheme calculations. No signature of an unusually small $H(q)$ was found, even for low-salt systems close to the freezing line, as reported by other authors[10, 11]. Extensive calculations of the peak values, $H(q_m)$, of the hydrodynamic function of fluid systems, characterized by the empiric criterion $S(q_m) < 3.1$, in a broad range of $\phi$ and $Z$ values have revealed a characteristic region where all the values for $H(q_m)$ can be found. This type of analysis can be of help in the interpretation of existing experimental data, and in identifying conditions where either hydrodynamic hindrance or enhancement of collective diffusion may be found in the investigated colloidal system. We note that the upper limiting contour line in Figs. 7b and 8 depends not critically on the peak value of the freezing criterion. Changing this value from 3.1 to 2.9, say, leads only to a small shift in this curve.

In the present study, only systems without noticeable indications of crystallization have been considered, and interpreted in terms of theoretical schemes designed for fluid systems. Partially and fully crystalline suspensions require a special analysis of the measured scattering functions by methods which account for the long-range particle ordering. Such an analysis for silica spheres in DMF is in progress. The results of this analysis will be communicated in a future publication.

In addition to the study of $H(q)$ and $S(q)$, we have also investigated the salt and colloid concentration dependence of the effective charge numbers, $Z_{RY}$ and $Z_{el}$, obtained, respectively, from the Rogers-Young static structure factor fit based on the OMF pair potential, and from single-particle electrophoresis measurements. A second branch of high-charge values for $Z_{RY}$ was found to exist which leads to equally good fits of the experimental data, but was ruled out for physical reasons. The general trends in the $C_s$ and ϕ-dependence of the physical branch values of $Z_{RY}$ conform to those of $Z_{el}$ and the renormalized particle charge obtained from the renormalized jellium model. However, the rise in $Z_{RY}$ with increasing amount of added salt is much steeper, an observation which we can not explain to date. A deeper understanding of these effective charges and their interrelations will require a more detailed theoretical and experimental analysis, involving the consideration of charge-regulation[58, 59] at the silica surfaces, and colloid correlation effects, in particular when the electrophoretic mobility is considered. Correlation effects are especially relevant for low-salt suspensions with strongly overlapping electric double layers. Notwithstanding the progress made very recently in our understanding of structural and electro-hydrodynamic effects in suspensions of interacting charged colloidal particles[56, 57, 60, 61, 62, 63], a lot more needs to be learned about the various types of static and dynamic effective charges. This is left to future work.

**Acknowledgments**

This work has been supported by the Deutsche Forschungsgemeinschaft (SFB-TR6, project section B2). Part of this work was done within the framework of the "SoftComp" Network of Excellence (No. S080118). AJB acknowledges financial support from FONCYT (OICT 2005-33691) and SeCyT-UNC.